\documentclass[preprint,12pt]{elsarticle}

\usepackage{graphics}
\usepackage{graphicx}
\usepackage{amsmath}
\usepackage{amsthm}
\usepackage{epstopdf}
\usepackage{dcolumn}% Align table columns on decimal point
\usepackage{bm}% bold math
\usepackage{color}
\usepackage[%showframe,%Uncomment any one of the following lines to test 
total={6.5in,8.75in}, top=1.5in, left=0.9in, %includefoot,
]{geometry}

\biboptions{comma,sort,compress}

\journal{Physica A}

\begin{document}

\begin{frontmatter}

%% Title, authors and addresses

%% use the tnoteref command within \title for footnotes;
%% use the tnotetext command for the associated footnote;
%% use the fnref command within \author or \address for footnotes;
%% use the fntext command for the associated footnote;
%% use the corref command within \author for corresponding author footnotes;
%% use the cortext command for the associated footnote;
%% use the ead command for the email address,
%% and the form \ead[url] for the home page:
%%
%% \title{Title\tnoteref{label1}}
%% \tnotetext[label1]{}
%% \author{Name\corref{cor1}\fnref{label2}}
%% \ead{email address}
%% \ead[url]{home page}
%% \fntext[label2]{}
%% \cortext[cor1]{}
%% \address{Address\fnref{label3}}
%% \fntext[label3]{}

\title{Income Distribution in the European Union Versus in the United States}

%% use optional labels to link authors explicitly to addresses:
%% \author[label1,label2]{<author name>}
%% \address[label1]{<address>}
%% \address[label2]{<address>}

\author{Maciej Jagielski\corref{cor1}}
\ead{zagielski@interia.pl}
\cortext[cor1]{Corresponding author. Tel.: +48 22 5532730; fax: +48 22 55 32 999.}
%\email{zagielski@interia.pl}
\author{Rafa\l{} Duczmal}
\author{Ryszard Kutner}%

\address{Faculty of Physics, University of Warsaw, Pasteura 5, PL-02093 Warszawa, Poland}

\begin{abstract}
We prove that the refined approach -- our extension of the Yakovenko et al. formalism -- is universal in the sense that it describes well both household incomes in the European Union and the individual incomes in the United States for social classes of any income. This formalism allowed the study of the impact of the recent world-wide financial crisis on the annual incomes of different social classes. Hence, we indicate the existence of a possible precursor of a market crisis. Besides, we find the most painful impact of the crisis on incomes of all social classes. 
\end{abstract}


\end{frontmatter}

%%
%% Start line numbering here if you want
%%
% \linenumbers

\section{Introduction}

One of the major trends having a long history in socio- and econophysics is the study of income and wealth redistribution in society and the analysis of social inequalities. Several models trying to explain the microscopic mechanisms of income dynamics of individuals or households were proposed \cite{RHCR_2006, YR_2009, BY_2010, G_1931, K_1945, A_1995, S_1997, SR_2001, RS_2001, SR_2002, H_2004}.

However, (to the best of our knowledge) none of the models that have been developed so far give an analytic description of the annual household or individual incomes of all social classes (i.e. the low-, medium-, and high-income classes) by a single formula based on a unified formalism. Recently \cite{JK_2013b}, we extended the Yakovenko et al. model providing, indeed, such a unified formalism. 

In the present paper we show that the formula which we derived within this unified formalism, containing a low number of free parameters,  satisfactorily reproduces the empirical complementary cumulative distribution functions (CCDFs) both for the European Union (EU) and for the United States (US). The cumulative distribution function \footnote{The complementary cumulative distribution function is the probability that the independent stochastic variable takes a value larger than some fixed one.} is the main statistical tool used in this context, that is, the descriptive statistics technique is involved herein to analyse data. 

\section{Comments on formula}

To describe the income of all social classes in the US and the EU, we used our extended Yakovenko et al. formalism (EYF) \citep{JK_2013a, JK_2013b}. 

As for the Yakovenko et al. model, the coexistence of multiplicative and additive processes on the level of the Langevin equation and hence the Fokker-Planck one, is also allowed for the EYF. That is, we assume that household or individual incomes are determined by: (i) systematic wages and salaries and/or (ii) random profits that go to households or individuals mainly through financial investments and/or capital gains. Furthermore, for the EYF we assume that the formalism of the income change is the same for the entire society, however, its detailed dynamics distinguishes well the ranges of individual income social classes, in particular, of the high-income social class from that of others (see \cite{JK_2013b} for details).

We found, in the framework of the EYF, the equilibrium probability distribution function in the form \citep{JK_2013a, JK_2013b}
\begin{eqnarray}
P_{\rm eq}(m) \propto \left\{ \begin{array}{ll}
\, \frac{\exp\left(-(m_0/T)\arctan(m/m_0)\right)}{[1+(m/m_0)^2]^{(\alpha +1)/2}}, & \textrm{if $m<m_1$} \\
\, \frac{\exp\left(-(m_0/T_1)\arctan(m/m_0)\right)}{[1+(m/m_0)^2]^{(\alpha_1 +1)/2}}, & \textrm{if $m\ge m_1$}
\end{array} \right.
\nonumber \\\label{rown2}
\end{eqnarray}
where parameter $m_0$ is a crossover (border) income between the low- and medium-income society classes, while parameter $m_1$ is an analogous border income but between the medium- and high-income social classes. Parameter $T$ can be interpreted as an average income per household or individual within the low- and medium-income social classes, while interpretation of parameter $T_1$ is given further in the text. The shape parameters $\alpha$ and $\alpha_1$ are the Pareto exponents, describing the income inequality within the medium- and high-income society classes, respectively. The CCDF considered below is, indeed, an integrated quantity of the above given distribution function.

\section{Remarks on databases}

In the case of the European Union we exploit the empirical data from Eurostat's Survey on Income and Living Conditions (EU-SILC) \cite{EURO_2005, EURO_2006, EURO_2007, EURO_2008, EURO_2009, EURO_2010} for the years 2005-2010. This database contains information on the demographic characteristics of households, their living conditions, as well income as economic activity. In our analysis we chose the \emph{total household gross income} variable. However, Eurostat's EU-SILC database contains only a few observations concerning the income of households belonging to the high-income social class, which is insufficient to subject to any statistical description. In order to improve the statistics for the high-income social class, we additionally analysed the effective income of billionaires \footnote{The term {\it billionaire} used herein is equivalent (as in the US terminology) to the term {\it multimillionaire} used in the European terminology.} in the EU by using the Forbes ranking `The World's Billionaires' \cite{FORBES} (see \citep{JK_2013a, JK_2013b} for more details). 

In the case of the United States we used the empirical data from the Internal Revenue Service (IRS), the US government tax agency, for the years 2005-2010 \citep{IRS}. We chose the \emph{adjusted gross income} variable as the only one accessible in the context of our comparative analysis. Similarly, as for the EU-SILC, the IRS database does not contain observations on the individuals belonging to the high-income social class. Again, in order to consider the high-income social class, we additionally analysed the effective income of billionaires in the US by using the same Forbes ranking as mentioned above.

By using the EU-SILC database as well as the rank of the richest Europeans and the IRS dataset and a ranking of the richest Americans, we were able to consider incomes of all social classes thanks to the joint procedure presented in details in Refs. \citep{JK_2013a, JK_2013b}. Thus, we obtained a data record sufficiently large for the statistical consideration of all social classes.

\section{Results and discussion}

We compared the theoretical CCDF, based on the probability distribution function $P_{\rm eq}(m)$ given by Eq. (\ref{rown2}), with: (i) the empirical CCDF of the annual total gross income of households in the EU and the corresponding (ii) empirical CCDF of the annual adjusted gross income of individuals in the US. In our studies we analysed the empirical CCDF constructed by using the well-known Weibull rank formula \citep{CMM_1988, Han_2004}.

\begin{figure}[ht]
\centering
\includegraphics[scale=0.40]{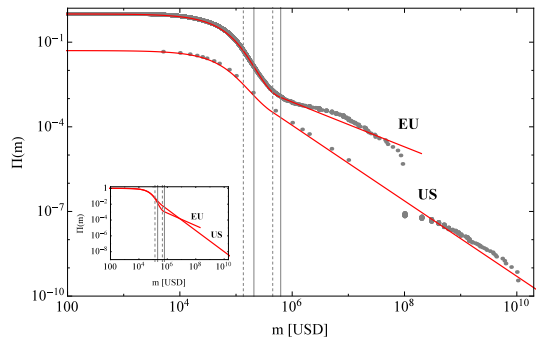}
\caption{A typical comparison of the theoretical CCDFs (solid curves) with the EU household income empirical data set (dots -- the upper curve) and the US individual income data set (dots -- the lower curve), for instance for the year 2007. Notably, the US (theoretical and empirical) curves were shifted down by about one and a half decade for better distinguishing -- their original location is shown in the miniature plot containing only the theoretical CCDFs. The solid and dashed pairs of vertical lines concern the EU and the US curves, respectively. For both pairs the first vertical line is placed at $m_0$, while the second one is at $m_1$. Apparently, the medium-income social class is much more distinctly formed for the EU than for the US.}
\label{fig2}
\end{figure}

\begin{figure}[ht]
\centering 
\includegraphics[scale=0.40]{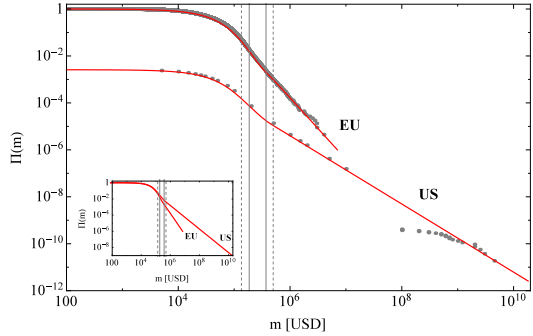}
\caption{A comparison of the theoretical CCDFs (solid curves) with the EU household income empirical data set (dots -- the upper curve) and the US individual income data set (dots -- the lower curve) for the exceptional year 2009. For better distinguishing, the US curves were shifted down by about two and a half decade -- their original location is shown in the miniature plot containing only the theoretical CCDFs. The solid and dashed pairs of vertical lines play the same role as in Fig. \ref{fig2}. It is striking that the high-income social class almost disappeared in the EU in comparison with the US -- for better verification see Fig. \ref{fig2} which also shows how stable the US curve is.}
\label{fig3}
\end{figure}

The two resulting plots, each consisting of the theoretical (solid curves) and empirical (small circles) CCDFs for the EU (the upper curves) and the US (the lower curves), are presented in a log-log scale in Figs. \ref{fig2} and \ref{fig3} for a typical year, 2007, and an exceptional year, 2009. Apparently, the EYF describes both the EU and the US empirical CCDFs well. Hence, we were able to provide estimates of the EYF parameters for the years 2005-2010, both for European Union households and United States individuals (cf. Tables \ref{tab2} and \ref{tab3}). Notably, fits were the best for $T_1=m_1$, which also gives the interpretation of parameter $T_1$.

\begin{table}[ht]
\centering
\caption{Parameters $m_0$, $T$, and $T_1(=m_1)$ obtained, in US dollars, for the years 2005-2010, from the comparison of the theoretical CCDF with the corresponding
empirical ones concerning the annual: (i) total gross income of households in the EU and (ii) adjusted gross income of individuals in the US. The error bars of the parameters do not exceed $18\%$.}
 \begin{tabular}{|c|c|c|c||c|c|c|}
  \hline
 & \multicolumn{3}{|c||}{\bf European} & \multicolumn{3}{|c|}{\bf United} \\
 & \multicolumn{3}{|c||}{\bf Union} & \multicolumn{3}{|c|}{\bf States} \\
  \hline 
 {\bf Year} & $\boldsymbol{m_0}$ & $\boldsymbol{T}$ & $\boldsymbol{T_1}$ & $\boldsymbol{m_0}$ & $\boldsymbol{T}$ & $\boldsymbol{T_1}$ \\
\hline 
	2005 & $199\,254$ & $46\,278$ & $552\,770$ & $135\,000$ & $45\,520$ & $380\,000$ \\
  \hline 
	2006 & $172\,373$ & $43\,985$ & $529\,006$ & $150\,000$ & $47\,220$ & $350\,000$ \\
  \hline
	2007 & $208\,116$ & $48\,127$ & $624\,350$ & $135\,000$ & $48\,430$ & $450\,000$ \\
  \hline
	2008 & $174\,495$ & $55\,257$ & $654\,355$ & $135\,000$ & $48\,740$ & $460\,000$ \\
  \hline
	2009 & $185\,945$ & $47\,448$ & $371\,890$ & $135\,000$ & $48\,050$ & $500\,000$ \\
  \hline
	2010 & $183\,225$ & $51\,574$ & $610\,749$ & $135\,000$ & $48\,680$ & $420\,000$ \\
	\hline
\end{tabular} 
\label{tab2}
\end{table}

\begin{table}[ht]
\centering
\caption{Exponents $\alpha$ and $\alpha_1$ obtained for the years 2005-2010 from the comparison of the theoretical CCDF with: (i) the empirical CCDF of the annual total gross income of households in the EU, and (ii) the empirical CCDF of the annual adjusted gross income of individuals in the US. The error bars of exponents do not exceed $4\%$.}
\begin{tabular}{|c|c|c||c|c|}
  \hline
 & \multicolumn{2}{|c||}{ \bf European} & \multicolumn{2}{|c|}{ \bf United}\\
 & \multicolumn{2}{|c||}{ \bf Union} & \multicolumn{2}{|c|}{ \bf States}\\
  \hline 
 {\bf Year} & $\boldsymbol{\alpha}$ & $\boldsymbol{\alpha_1}$ & $\boldsymbol{\alpha}$ & $\boldsymbol{\alpha_1}$ \\
\hline
	2005 & $2.907$ & $0.795$ & $1.93$ & $1.354$ \\
  \hline
	2006 & $2.892$ & $0.86$ & $1.88$ &$1.346$ \\
  \hline
	2007 & $2.735$ & $0.79$ & $1.83$ & $1.336$ \\
  \hline
	2008 & $2.965$ & $0.890$ & $1.85$ & $1.381$ \\
  \hline
	2009 & $2.974$ & $2.608$ & $1.90$ &$1.451$ \\
  \hline
	2010 & $3.153$ & $0.77$ & $1.86$ & $1.395$ \\
	\hline
\end{tabular} 
\label{tab3}
\end{table}

Remarkably, the values of borders $m_0$ and $m_1$ are systematically larger for the EU than for the US, except for 2009 (this meaningful exception is discussed further in the text). The systematic deviation is mainly caused by the fact that we compare the household incomes \footnote{In average, there are about 1.5 employers per single EU household.} in the EU with individual incomes in the US.

Apparently, the range of a medium-income social class (equal to $m_1-m_0$) is reduced (typically by about 15\%) in the case of the US in comparison with the EU (cf. Fig. \ref{fig2}). This is a persistent result except for the year 2009, i.e. valid for almost every considered year (in our case from 2005 to 2010). The medium-income social class is more distinct in the EU than in the US mainly because the difference CCDF($m_0$)-CCDF($m_1$) is greater for the EU (by a factor of about 1.5). This estimation is also confirmed by the slopes of CCDFs for the EU and the US -- the ratio of both slopes again gives a value equal to about 1.5 -- for the verification, the corresponding Pareto exponents (given in Table \ref{tab3}) can be compared (see also Fig. \ref{fig2}). 

Besides, the border $m_0$ increased at the very beginning of the recent world-wide financial crisis by about 10\% -- in 2006 in the US and in 2007 in the EU (cf. Table  \ref{tab2}) -- increasing, thereby, the ranges of the corresponding low-income social classes. Although later these borders returned to their typical values, this effect could be identified as a clearly interpretable possible early-warning signal preceding the crisis -- however, to say something more definitive, a comparative systematic study concerning all other crises is required.  

Although the border $m_1$ in the EU during the exceptional year 2009 was decreased by a factor of about 1.7 in comparison with its typical value (see Table \ref{tab2}), the range of the high-income social class decreased. This is because the upper limit of this class drastically decreased by more than one decade (compare the upper curves in Figs. \ref{fig2} and \ref{fig3}). Furthermore, since the border $m_0$ practically did not change (cf. Table \ref{tab2}), the resulting distinct shrink of the medium-income social class is observed. 

These observations directly relate to the most striking observation that the high-income social class is driven, in the EU, by exponent $\alpha_1$ which during 2009 almost equals $\alpha$ (up to about 10\% accuracy, see Table \ref{tab3}). This observation leads to a situation where the high-income social class plays the role of the medium-income one (see Fig. \ref{fig3} for details). Hence, although exponent $\alpha _1$ returned later on to its typical value, this result could be identified as a particularly drastic impact of the crisis. That is, the high-incomes social class almost disappeared in the EU (cf. Fig. \ref{fig3}). Even though for the same year in the US we also observed a slight increase of exponent $\alpha_1$, its values are still lower than $\alpha$ (by about 25\%). Hence, the shape of the CCDF is quite stable for the US for the years 2005-2010, in spite of a remarkable increase of the range of the medium-incomes social class (by about 10\%) in 2009. 

\section{Conclusions}

In the present paper we demonstrated (to best of our knowledge) the first comparison of incomes in the EU and the US done in such a systematic way. It was possible because we applied the extended Yakovenko et al. formalism. We proved, herein, that the EYF describes the income of the EU households and the US individuals well. By using the EYF we show that both in the EU and in the US we deal with three income social classes, where the medium-income social class has only an intermediate character -- one can even say that both in the US and in the exceptional year 2009 in the EU, it has a residual character. 

We found that in the year 2009 the high-income social class abruptly took the role of the medium-income one. This means that the high-income social class, in practice, vanished. That is, the EU society as a whole became poorer than during other years -- this is the most drastic impact of the crash. In contrast, in 2009 in the US, only a small increase of the medium-income social class was observed, making the range of the high-income social class shorter. Thus, we show that the crisis in the EU has a relatively more painful character than in the US. During the next year the situation returned to its typical state. 

Furthermore, an abrupt increase of the upper border of the low-income social class (in the year 2006 in the US and one year later in the EU) can be considered as an early-warning signal before the crisis. Nevertheless, the low-income social class in the EU is very similar to the corresponding one in the US -- the shape of both CCDFs is quite stable. The crisis was more painful for the medium- and high-income social classes than for the low-income one.

We can conclude that the complementary cumulative distribution function, although being a global (macroeconomic) characteristic, is sufficiently sensitive to the crises and crashes, clearly responding over the extended Yakovenko et al. formalism to the income situation in each income social class, at least in the EU and the US.

\section*{Acknowledgements}
We thank Victor M. Yakovenko for his helpful discussion.

\bibliographystyle{apsrev4-1}
\bibliography{MJ_RK_PhysA}

\end{document}